\documentclass[aps, prb, reprint,superscriptaddress]{revtex4-2}
\usepackage{graphicx}%
\usepackage{dcolumn}
\usepackage{bm}
\usepackage{amsmath,amssymb}
\usepackage{color} 

\begin{document}
\preprint{kkk}

\hyphenation{cryo-cooler}

\title{Insulating Phase in Two-dimensional Josephson-Junction Arrays Investigated by Nonlinear Transport}



\author{Hiroki~Ikegami}
\altaffiliation{Present address: Institute of Physics, Chinese Academy of Sciences} 
\email{hikegami@iphy.ac.cn}
\affiliation{RIKEN Center for Quantum Computing (RQC), Wako, Saitama 351-0198, Japan}

\author{Yasunobu~Nakamura}
\affiliation{RIKEN Center for Quantum Computing (RQC), Wako, Saitama 351-0198, Japan}
\affiliation{Department of Applied Physics, Graduate School of Engineering, The University of Tokyo, Bunkyo-ku, Tokyo 113-8656, Japan}

\date{\today}

\begin{abstract}

We present experimental investigations of transport properties in the insulating phase of two-dimensional Josephson-junction arrays (JJAs) by systematically changing the ratio of Josephson energy $E_\mathrm{J}$ and charging energy $E_\mathrm{C}$. 
The observed temperature dependence of resistance indicates that the JJAs do not show a sharp phase transition but exhibit a gradual crossover to the insulating phase. 
At low temperatures, the current--voltage ($I$--$V$) characteristics become nonlinear as described by $I=cV+bV^a$ ($a$, $b$, and $c$ are temperature dependent coefficients).  
This nonlinear behavior is understood in terms of the Berezinskii-Kosterlitz-Thouless (BKT) mechanism by taking into account the influence of a finite-range cutoff of the logarithmic interaction between Cooper pairs.  
From the analysis of the nonlinearity, we deduce the crossover temperature to the insulating phase and determine the phase diagram in the insulating side as a function of $E_\mathrm{J} /E_\mathrm{C}$.  
We also show that, at very low temperatures, the $I$--$V$ characteristics continuously develop into the negative differential conductance caused by coherent single-Cooper-pair tunneling.

\end{abstract}


\maketitle

\section{Introduction}
The superconductor-insulator transition (SIT) that occurs at zero temperature is a representative example of quantum phase transitions (QPTs)~\cite{Goldman_IntJModPhysB2010,Sondhi_RMP1997,Gantmakher_PhysUsp2010}.  
It is observed in a wide range of systems from amorphous metal films~\cite{Haviland_PRL1989,Markovic_PRL1998,Baturina_PRL2007,Ovadia_NatPhys2013} to superconducting oxides~\cite{Bollinger_Nature2011}.
In some of the systems, the insulating phase occurs not by breaking Cooper pairs but by localization of Cooper pairs~\cite{Goldman_IntJModPhysB2010,Markovic_PRL1998,Baturina_PRL2007,Bollinger_Nature2011,Fisher_PRL1990-1,Fisher_PRB1989}.
The SIT to such a Bose insulating phase involves only bosonic degrees of freedom, and investigations of such a simple system offers a starting point to explore more complex QPTs.
To understand the nature of the QPT to the Bose insulating phase, uncovering detailed properties of the insulating phase is of prime importance.

A pivotal model system to gain a deeper understanding of the Bose insulating phase is Josephson-junction arrays (JJAs)~\cite{Fazio_PhysRep2001,Geerligs_PRL1989,Mooij_PRL1990,vanderZant_PRL1992,vanderZant_PRB1996,Tighe_PRB1993,Delsing_PRB1994,vanOudenaarden_PRL1996,vanOudenaarden_PRB1998,Yamaguchi_PRB2006,Cosmic_PRB2018,Kuzmin_NatPhys2019}.
They are artificial quantum many-body systems composed of superconducting islands connected via small Josephson junctions. 
In two-dimensional (2D) JJAs, the QPT between the superconducting and insulating phases occurs due to the competition between two energies: the Josephson energy $E_\mathrm{J}$, which allows for tunneling of Cooper pairs to neighboring islands, and the charging energy $E_\mathrm{C}$, which tends to pin Cooper pairs to each island~\cite{Geerligs_PRL1989,Mooij_PRL1990,vanderZant_PRB1996,Fazio_PRB1991,Fazio_PhysRep2001}.
Intensive experimental studies have clarified fundamental properties in the superconducting phase realized at $E_\mathrm{J} \gg E_\mathrm{C}$, such as the Berezinskii-Kosterlitz-Thouless (BKT) transition to superconducting phase~\cite{Resnick_PRL1981,Abraham_PRB1982,vanWees_PRB1987,Cosmic_PRB2020,Nerwock_Book} and the phase diagram of the superconducting phase as a function of $E_\mathrm{J}/ E_\mathrm{C}$~\cite{vanderZant_PRB1996}.
Compared to the superconducting phase, however, nature of the insulating phase realized at $E_\mathrm{C} \gg E_\mathrm{J}$ has been less investigated experimentally.

A key concept that may illuminate properties of the insulating phase is duality, which is approximately present between vortices and charges in 2D JJAs~\cite{Fazio_PRB1991,Fazio_PhysRep2001}.
From the duality argument, the thermodynamic transition to the insulating phase is driven by the BKT mechanism of binding of Cooper pairs and anti-Cooper pairs (charge BKT transition).
This is associated with the fact that Cooper pairs and anti-Cooper pairs, which are elementary excitations in the insulating phase, interact with the logarithmic potential as analogous to vortices in the superconducting phase, if screening of the interaction by the capacitance to ground is neglected (see Sect.~II).
However, divergence of the resistance at a nonzero temperature expected from the charge BKT transition has not been observed experimentally~\cite{Mooij_PRL1990,Tighe_PRB1993,Delsing_PRB1994}.
This may be due partly to smearing of the transition by screening of the interaction by the capacitance to ground.
Kanda \textit{et al}.\ discussed their transport data in terms of the charge BKT transition by taking into account the influence of the screening~\cite{Kanda_JSPSJ1994,Kanda_JSPSJ1995-2} although the investigated range of $E_\mathrm{J} / E_\mathrm{C}$ was limited.

In this article, we present systematic investigations, in particular, of nonlinear transport over a wide range of $E_\mathrm{J}/E_\mathrm{C}$ and discuss how the insulating phase is formed when the temperature is decreased. 
Investigation of nonlinear transport is the key to understanding properties of the insulating phase because a specific power-law dependence of current--voltage ($I$--$V$) characteristics as well as the universal jump in the power exponent are expected in the charge BKT transition~\cite{Fazio_PRB1991}.
We also show that, at low temperatures, the $I$--$V$ characteristics continuously develop into the negative differential conductance arising from  coherent single-Cooper-pair tunneling.

\section{JJA and charge BKT transition}

In a JJA, each superconducting island has two quantum variables: the order parameter phase $\phi_j$ and the number of excess Cooper pairs $n_j$ (here the subscript $j$ is a label of the island).
These are conjugate variables satisfying the commutation relation $[\phi_j, n_k]=i\delta_{j,k}$. 
Using these variables, the Hamiltonian of the JJA in a magnetic field is described by
\begin{equation}
H = \frac{(2e)^2}{2} \sum\limits_{\left\langle {i,j} \right\rangle} n_i C^{-1}_{ij} n_j - E_\mathrm{J}\sum\limits_{\left\langle {i,j} \right\rangle} \cos(\phi_i-\phi_j -A_{ij}) ,
\label{eq:Hamiltonian}
\end{equation}
where $e$ is the elementary charge, $C_{ij}$ is a capacitance matrix element composed of the capacitance between nearest-neighbor islands, $C_\mathrm{J}$, and that between each island and the ground, $C_\mathrm{g}$, and $A_{ij}=(2\pi /\Phi_0) \int_i^j {{\bf{A}} \cdot d\mbox{\boldmath $l$}} $ is the line integral of the vector potential $\bf{A}$ from an island $i$ to an island $j$ with flux quantum $\Phi_0=h/2e$ ($h$ is the Planck constant)~\cite{Fazio_PhysRep2001}.
The first term is associated with the charging effect with an energy scale of $E_\mathrm{C}=e^2/(2C_\mathrm{J})$, and the second term represents the Josephson effect characterized by the Josephson energy $E_\mathrm{J}$.
Below we consider the case in zero magnetic field unless otherwise mentioned.
At $E_\mathrm{J} \gg E_\mathrm{C}$, the JJA exhibits the superconducting transition with decreasing temperature.
The transition is driven by binding of vortex and anti-vortex pairs associated with the BKT mechanism.
At $E_\mathrm{C} \gg E_\mathrm{J}$, on the other hand, the charging effect prevents tunneling of a Cooper pair to the neighboring island because of the Coulomb blockade, resulting in the insulating phase at ground state.
Because of the competition between the Josephson effect and the charging effect, the quantum phase transition between the superconducting and insulating phases occurs at  $E_\mathrm{J} \sim E_\mathrm{C}$~\cite{Fazio_PRB1991,Fazio_PhysRep2001,vanderZant_PRB1996} [see the phase diagram shown in Fig.~\ref{Fig2}(d)].
We note that, even in the insulating phase, Cooper pairs exist in each island.
However, the phase coherence over the system is destroyed because of the large fluctuations of the phase under the fixed number of Cooper pairs in each island.

At $E_\mathrm{C} \gg E_\mathrm{J}$, fluctuations of $\left\{ n_i \right\}$ are suppressed.
Thus, the JJA is characterized by a well-defined set of $\left\{ n_i \right\}$ across the whole array.
Under this situation, we consider an isolated charge $-2e$ added on an island.
This charge generates polarization on the surrounding islands within the spatial range of $\Lambda = \sqrt{C_\mathrm{J} /C_\mathrm{g}}$ unit cells.
Such a Cooper pair dressed with the polarization is referred to as a Cooper-pair soliton~\cite{Fazio_PRB1991,Mooij_PRL1990}.
The charge also produces an electrostatic potential $V(r)=-e/(\pi C_\mathrm{J}) K_0 (r/\Lambda)$ at distance $r$, where $K_0 (x)$ is the modified Bessel function.
In the limit of $r/\Lambda \ll 1$, this potential is approximated to be $V(r)=e/(\pi C_\mathrm{J}) \ln (r/\Lambda)$.
Thus, in the case of $ C_\mathrm{g}=0$, the interaction between a pair of charges $+2e$ and $-2e$ with a separation $r$ is given by~\cite{Fazio_PRB1991,Mooij_PRL1990}
\begin{equation}
U_\mathrm{p} (r) = \frac{4E_\mathrm{C}}{\pi} \ln (r).
\label{eq:log-interaction}
\end{equation}
The system is therefore described as a 2D Coulomb gas with the logarithmic interaction, where the BKT transition to the insulating phase is expected at low temperatures~\cite{Berezinskii_SovPhysJETP1971,Kosterlitz_JPhysC1973}.
The transition from the normal to insulating phases with decreasing temperature is driven by the BKT mechanism associated with binding of a Cooper pair and an anti-Cooper pair.
In the limit of $E_\mathrm{J}=0$, the transition temperature is $T^\mathrm{BKT}_\mathrm{i}=E_\mathrm{C}/ (\pi \varepsilon  k_\mathrm{B})$ \cite{Fazio_PRB1991,Mooij_PRL1990}, where $\varepsilon$ is a non-universal constant of an order unity and $k_\mathrm{B}$ is the Boltzmann constant.
We note that there is an approximate duality between the charges and vortices~\cite{Fazio_PRB1991,Fazio_PhysRep2001}.
Thus, as mentioned above, the transitions at $E_\mathrm{C} \gg E_\mathrm{J}$ and $E_\mathrm{J} \gg E_\mathrm{C}$ are both driven by the BKT mechanism mediated by binding of charge and vortex pairs, respectively.

In the case of $C_\mathrm{g} \neq 0$, the ground capacitance screens the logarithmic interaction at $r > \Lambda$~\cite{Mooij_PRL1990,Fazio_PhysRep2001,Bradley_PRB1984}.
Then, $U_\mathrm{p} (r)$ falls off exponentially at $r \gg \Lambda$, while $U_\mathrm{p} (r)$ is still logarithmic at $r \ll \Lambda$.
This suggests that the BKT mechanism works only at $r < \Lambda$.
Hence, charges with a separation shorter than $\Lambda$ can form bound pair while those separated further than $\Lambda$ cannot. 
This fact implies that the system size is effectively finite with a size of $\sim \Lambda$, causing smearing of the transition and modification of properties of the transition.

\section{Experimental}

\begin{figure}
	\begin{center}
		\includegraphics[keepaspectratio]{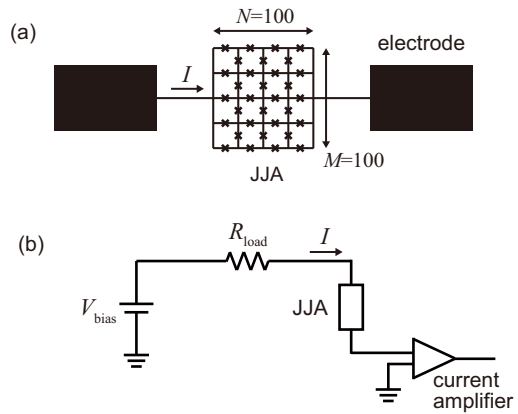}
	\end{center}
	\caption{\label{Fig1} 
	(a) Josephson junction array~(JJA). It consists of 100$\times$100 plaquettes connected to two electrodes. Islands at right and left edges are galvanically connected without junctions. (b) Schematic diagram of transport experiments employed in the insulating regime. Bias voltage $V_{\rm bias}$ is applied on a JJA through a load resistance $R_\mathrm{load}$, and current $I$  is detected. Voltage $V$ on the JJA is determined from $V=V_{\rm bias}-R_\mathrm{load}I$.
	}
\end{figure}

Our JJAs consist of superconducting islands of Al films connected in a square network of 100$\times$100 plaquettes via Josephson junctions [Fig.~\ref{Fig1}(a)], fabricated on top of a silicon substrate. 
The size of each Josephson junction is 100$\times$200~nm$^2$ and the area of each plaquette is 6$\times$6~$\mu$m$^2$.
The JJAs have two different kinds of edges: Islands are galvanically connected without junctions at the left and right edges which are connected to respective electrodes, while islands at the top and bottom edges are connected via Josephson junction.
The Josephson energy $E_\mathrm{J}$ is determined from normal-state resistance of the junction at low temperatures using the Ambegaokar-Baratoff relation~\cite{Ambegaokar_PRL1963}.
The nearest neighbor capacitance $C_\mathrm{J}=$~1.7~fF (and thus $E_\mathrm{C}/k_\mathrm{B}=$~530~mK) is determined from the offset voltage $V_{\rm off}$ in the nonlinear current-voltage characteristics observed at high voltages $V \gg 2N\Delta_\mathrm{s} /e$ using the relation $V_{\rm off} = Ne/(2C_\mathrm{J})$~\cite{Chen_PRB1996,Tighe_PRB1993,Geerligs_PRL1989,Fazio_PhysRep2001}, where $\Delta_\mathrm{s}$ is the superconducting gap of Al films and $N=$~100 is the number of junctions in series.
The capacitance to ground, $C_\mathrm{g}$, is estimated to be 14~aF from a finite element calculation.
The screening length is thus $\Lambda=\sqrt{C_\mathrm{J} /C_\mathrm{g}}\sim$~11.
Inhomogeneity in $E_\mathrm{J}$ and $E_\mathrm{C}$ within a single array is estimated to be about 5\%.

Experiments are carried out using a dilution refrigerator.
DC transport properties are measured by a standard four-probe technique for superconducting JJAs.
For insulating JJAs, a two-probe configuration is employed, where the sample is biased with voltage $V_{\rm bias}$ through a load resistor $R_{\rm bias}$ and current $I$ is detected as shown in Fig.~\ref{Fig1}(b).
(Unless otherwise mentioned, $R_\mathrm{load}=$~1.2$\times$10$^{5}$~$\Omega$, which is the sum of resistances of electrical wires and filters in the measurement lines.)
The voltage on the sample, $V$, is determined from $V=V_{\rm bias}-R_\mathrm{load}I$.
All electrical wires inside the cryostat are carefully filtered; each wire has a $\pi$-filter with a cutoff frequency of 3~MHz at room temperature, RC low-pass filters with a cutoff frequency of 2.4 and 38~kHz at the 4-K and mixing-chamber plates, respectively, and a microwave filter at the base temperature.
A resistance of 19~k$\Omega$ is also inserted near the sample in each wire.
The JJA is placed in a very low magnetic field environment of $\sim$~6~$\times$10$^{-4}$ Gauss (i.e., $\sim$~1~$\times$10$^{-3}\Phi_0$ per plaquette) realized by using a $\mu$-metal and a superconducting shield as well as applying a cancellation field.

\begin{figure*}
	\begin{center}
		\includegraphics[keepaspectratio]{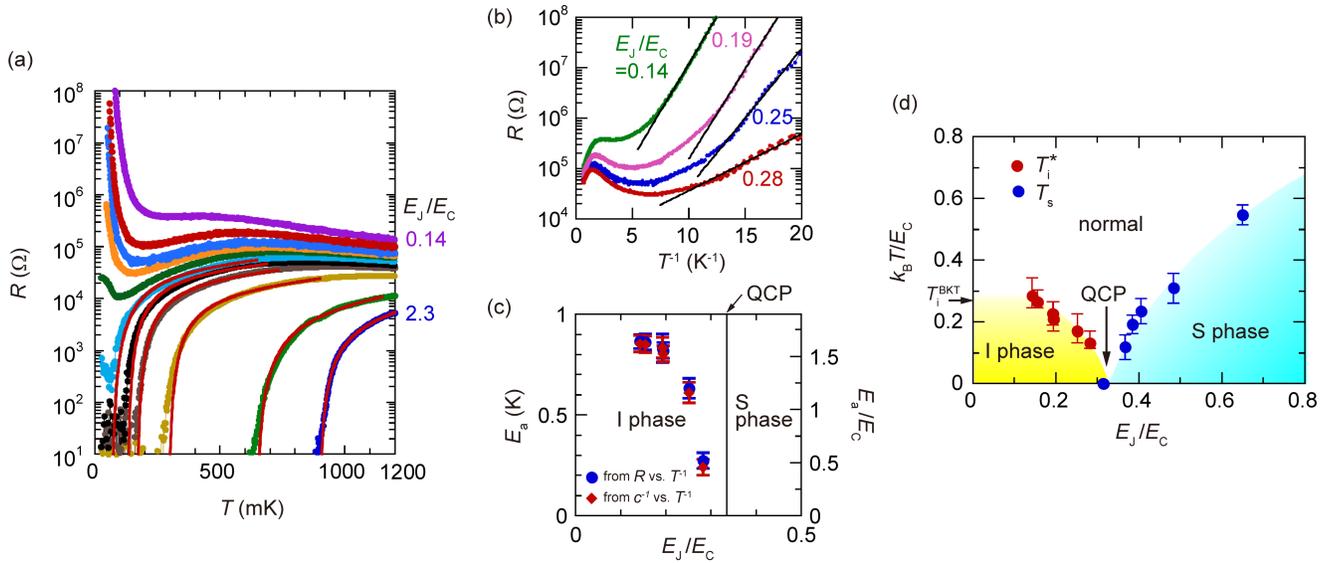}
	\end{center}
	\caption{\label{Fig2} 
		(a) Temperature dependence of the resistance of JJAs. From top to bottom, $E_\mathrm{J}/E_\mathrm{C}=$~0.14, 0.19, 0.25, 0.28, 0.31, 0.37, 0.41, 0.48, 0.65, 1.4, and 2.3. Red lines are the fitting to the BKT transition in the superconducting phase (see text). (b)~Resistance as a function of $T^{-1}$ in the insulating phase. Solid lines are the fitting to $R \propto \exp (E_\mathrm{a}/k_\mathrm{B}T)$. (c)~Activation energy $E_\mathrm{a}$ as a function of $E_\mathrm{J}/E_\mathrm{C}$ obtained from $R$ vs.\ $T^{-1}$ in (b) and $c^{-1}$ vs.\ $T^{-1}$ in Figs.~\ref{I-V_curve}(d)--(f), respectively. The vertical line indicates the quantum critical point (QCP) separating the superconducting (S) and  insulating (I) phases.  (d)~Phase diagram. The superconducting transition temperature $T_\mathrm{s}$ is determined by fitting to the expectation from the BKT theory. The crossover temperature $T_\mathrm{i}^\mathrm{*}$ is deduced from the nonlinearity observed in the $I$--$V$ characteristics (see text). The horizontal arrow indicates the theoretical charge BKT transition temperature $T^\mathrm{BKT}_\mathrm{i}$ in the limit of $E_\mathrm{J}=0$. The error bars in $T_\mathrm{s}$ are associated with the uncertainty depending on the choice of the temperature range for the fitting. The error bars in $T_\mathrm{i}^\mathrm{*}$ indicate the temperature region where the criterion of $a=3$ is within the error bars in $a$.
		}
\end{figure*}

\section{Formation of insulating phase}
\label{sect:insulating_phase}
In this section, we discuss how the insulating phase develops when the temperature is lowered, based on observed transport properties.
Figure~\ref{Fig2}(a) shows the temperature dependence of the zero-bias resistance $R$ measured by applying a small enough voltage to avoid nonlinearity.
At $E_\mathrm{J}/E_\mathrm{C} > 0.31$, the JJAs undergo the superconducting transition, while those with $E_\mathrm{J}/E_\mathrm{C} < 0.31$ exhibit the insulating behavior with diverging resistance at low temperatures.
Therefore, the JJA shows a quantum phase transition between the superconducting and insulating phases as $E_\mathrm{J}/E_\mathrm{C}$ is varied, as illustrated in the phase diagram in Fig.~\ref{Fig2}(d).
In the superconducting phase, the BKT transition temperature $T_\mathrm{s}$ was determined by fitting the temperature dependence of resistance  to $R/R_\mathrm{N} =C \exp\left( -B/\sqrt{t-t_\mathrm{s}} \right) $ with fitting parameters $B$, $C$, and $t_\mathrm{s}$ (solid lines in [Fig.~\ref{Fig2}(a)])~\cite{Lobb_PRB1983}, where $t=k_\mathrm{B}T/E_\mathrm{J}$, $t_\mathrm{s}=k_\mathrm{B}T_\mathrm{s}/E_\mathrm{J}$, and $R_\mathrm{N}$ is the resistance of the normal-state array.
In the insulating phase, a crossover between the normal and insulating phases occurs instead of a sharp transition as discussed later, and the crossover temperature $T_\mathrm{i}^\mathrm{*}$ determined as shown later is plotted in the phase diagram.
Hereafter, we focus on behaviors of the insulating phase.

The resistance in the insulating phase does not show divergence at a nonzero temperature, but it rather shows the Arrhenius behavior $R \propto \exp (E_\mathrm{a}/k_\mathrm{B}T)$ at low temperatures [Fig.~\ref{Fig2}(b)].
The activation energy $E_\mathrm{a}$ obtained from the fitting is on the order of $E_\mathrm{C}$ for small $E_\mathrm{J}/E_\mathrm{C}$ ($E_\mathrm{a} \sim k_\mathrm{B} \times 860$~mK $=1.6E_\mathrm{C}$ for $E_\mathrm{J}/E_\mathrm{C}= 0.14$) and decreases steeply toward the quantum critical point (QCP) as shown in Fig.~\ref{Fig2}(c).
The similar Arrhenius behavior has been observed in previous experiments~\cite{Tighe_PRB1993,Delsing_PRB1994,Kanda_JSPSJ1994,Kanda_JSPSJ1995-2}. 
According to Delsing \textit{et al}.\ \cite{Delsing_PRB1994}, there are two different mechanisms that generate the Arrhenius behavior depending on the ratio $E_\mathrm{C} /\Delta_\mathrm{s}$ ($\Delta_\mathrm{s}$ is the BCS gap of an Al film).
The first mechanism is thermal hopping of Cooper pairs, which occurs if $E_\mathrm{C}$ is small enough compared to $\Delta_\mathrm{s}$ (theoretically $E_\mathrm{C} /\Delta_\mathrm{s} < 4/3$).
This mechanism gives rise to $E_\mathrm{a}\sim E_\mathrm{C}$ when $E_\mathrm{J}$ is negligible.
On the other hand, if $E_\mathrm{C}$ is larger than $\Delta_\mathrm{s}$ (theoretically $E_\mathrm{C} /\Delta_\mathrm{s} > 4/3$), another mechanism sets in.
In this case, the activation energy can be decreased by breaking a Cooper pair into two quasiparticles on tunneling.
The activation energy in this case is given by $E_\mathrm{a}=\Delta_\mathrm{s}+\frac{1}{4}E_\mathrm{C}$ if $E_\mathrm{J}$ is negligible.
Our JJAs have $E_\mathrm{C}$ much smaller than $\Delta_\mathrm{s}$ ($E_\mathrm{C}/ \Delta _s\sim0.2$), indicating that hopping of Cooper pairs without pair breaking is responsible for the Arrhenius-type transport at low temperatures.

\begin{figure*}
	\begin{center}
				\includegraphics[keepaspectratio]{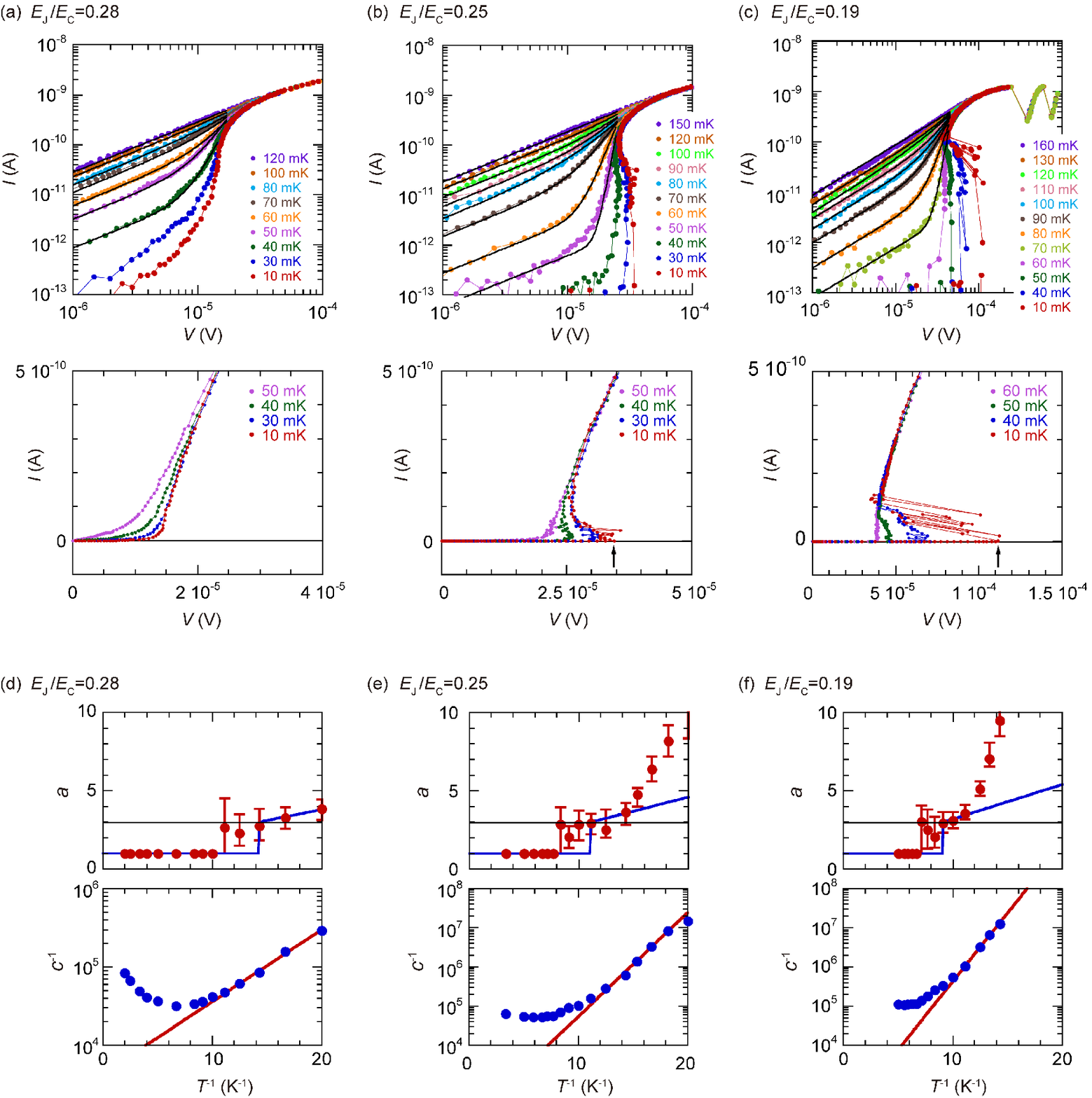}
	\end{center}
	\caption{\label{I-V_curve} 
		 (a)--(c) $I$--$V$ characteristics in the log scale (upper panel) and linear scale (lower panel). (a) $E_\mathrm{J}/E_\mathrm{C} =$~0.28, (b) 0.25, and (c) 0.19. Solid curves in upper panels are the fitting to $I=cV+bV^a$. The fittings are performed in a voltage range of less than  (a) 1.5$\times$10$^{-5}$~V, (b) 2.5$\times$10$^{-5}$~V, and (c) 4.5$\times$10$^{-5}$~V. Arrows in lower panels indicate the maximum voltage observed in the Coulomb-blockade region, $V_\mathrm{b}$. Data below (b)~50~mK and (c)~60~mK are measured with $R_\mathrm{load}=$ 1.12$\times$10$^6$~$\Omega$, and other data are taken with $R_\mathrm{load}=$ 1.2$\times$10$^5$~$\Omega$. (d)--(f) $a$~(upper panel) and $c^{-1}$~(lower panel) obtained from the fitting as a function of $T^{-1}$ for (d) $E_\mathrm{J}/E_\mathrm{C} =$~0.28, (e) 0.25, and (f) 0.19. The error bars in $a$ are associated with the uncertainty depending on the choice of the voltage range for the fitting. The solid blue curves in upper panels are the behavior expected from the charge BKT mechanism for $C_\mathrm{g}=0$ ($a=2T^\mathrm{BKT}_\mathrm{i}/T +1$ at $T<T^\mathrm{BKT}_\mathrm{i}$). The red solid lines in lower panels are the fitting to $c^{-1} \propto \exp (E_\mathrm{a}/k_\mathrm{B}T)$.
	}
\end{figure*}

clarify properties of the insulating phase in more detail, we investigate nonlinear transport phenomena.
As shown in Figs.~\ref{I-V_curve}(a)--(c), the $I$--$V$ characteristics exhibit notable nonlinear behaviors at low temperatures.
First, we discuss the data of $E_\mathrm{J}/E_\mathrm{C} =$~0.25 [Fig.~\ref{I-V_curve}(b)] as a typical example.  
At high temperatures ($T \gtrsim 150$~mK), $I$ is proportional to $V$ up to $\sim$~3$\times$10$^{-5}$ V [see upper panel of Fig.~\ref{I-V_curve}(b)].
As the temperature is lowered, nonlinearity emerges: The current shows a steeper increase around 2$\times$10$^{-5}$ V although it is proportional to $V$ at the low voltage region.
The increase around 2$\times$10$^{-5}$ V becomes further steeper at lower temperatures.
At temperatures below 40~mK, back-bending characterized by negative differential conductance ($dI/dV<0$) develops, which is more clearly seen in the linear plot shown in the lower panel of Fig.~\ref{I-V_curve}(b). 
We note that multiple jumps are observed at very high voltages ($V>$~2$\times$10$^{-4}$ V) [see upper panel of Fig.~\ref{I-V_curve}(c)], which are caused by successive row switching into a dissipative state~\cite{vanderZant_PRB1988}.
They are not a peculiar feature of the insulating phase but also observed in the superconducting phase.

In order to help understanding the observed behaviors, we first review nonlinear transport properties expected for the {\it unscreened} logarithmic interaction [i.e., Eq.~(\ref{eq:log-interaction}) expected for $C_\mathrm{g}=0$].
In this case, the charge BKT transition is expected to occur at the temperature $T^\mathrm{BKT}_\mathrm{i}$, and the $I$--$V$ characteristics should show a peculiar nonlinearity below $T^\mathrm{BKT}_\mathrm{i}$.
In particular, $I \propto V^a$ is expected, where $a=2T^\mathrm{BKT}_\mathrm{i}/T +1$, as a result of charge-pair breaking under large voltages~\cite{Fazio_PRB1991}.
(This $I$--$V$ relation can be derived in a way parallel to the relation of $V \propto I^a$ for the vortex BKT system \cite{Nerwock_Book,Halperin_JLTP1979,Lobb_PRB1983}.)
At $T > T^\mathrm{BKT}_\mathrm{i}$, on the other hand, $I$ is proportional to $V$ (i.e., $a=1$).
Therefore, the exponent $a$ shows a universal jump from 1 to 3 at $T^\mathrm{BKT}_\mathrm{i}$, which is a hallmark of the charge BKT transition.

In the case of $C_\mathrm{g} \neq 0$, as in our JJAs,  the logarithmic interaction is cut off at $r \sim \Lambda$ due to the screening by the ground capacitance.
At $r \gg \Lambda$, the interaction diminishes exponentially, and charges and anti-charges separated further than $\Lambda$ cannot form bound pairs even below the BKT temperature.
Therefore, the charge binding mechanism works only at $r < \Lambda$.
This suggests that the system size is effectively finite.
The BKT transition then becomes a continuous crossover rather than a sharp transition, resulting in rounding of the universal jump.
Furthermore, because Cooper pairs and anti-Cooper pairs separated further than $\Lambda$ cannot form bound pairs, free Cooper pairs exist even below $T^\mathrm{BKT}_\mathrm{i}$ with a density proportional to $\exp[-U_\mathrm{p}(\Lambda)/2k_\mathrm{B}T]$.
These free Cooper pairs experience dissipative transport under a small bias voltage, showing linear conductance.
Thus, $I=c V+bV^a$ with a smeared universality jump in $a$ is expected for $C_\mathrm{g} \neq 0$; the first term arises from transport of free Cooper pairs, and the second term represents nonlinearity associated with charge pairs bounded by the BKT mechanism.
This $I$--$V$ relation is indeed parallel with the nonlinear transport property $V=c I+bI^a$ in a finite-size vortex BKT system~\cite{Herbert_PRB1998}, where the first term arises from dissipative transport of free vortices present due to the finite size effect and the second term represents the nonlinear transport associated with the vortex BKT mechanism.

Based on the above argument, we fit the nonlinear data to $I=c V+bV^a$.
The data are well fitted as shown in Figs.~\ref{I-V_curve}(a)--(c) when the temperature is not too low.
The obtained exponent $a$ and the linear coefficient $c$ are shown in Figs.~\ref{I-V_curve}(e)--(f) as a function of $T^{-1}$.
With decreasing $T$, the exponent $a$ begins to increase from 1, where the $I$--$V$ characteristic deviates from the linear behavior.
We note that $a$ has large uncertainties around the temperature at which $a$ deviates from 1 because the linear term $c V$ dominates over the term $bV^a$ around this temperature.

The exponent $a$ is expected to show a universal jump from 1 to 3 at the transition in the case of $C_\mathrm{g}=0$.
For $C_\mathrm{g} \neq 0$, the transition becomes crossover and the jump should be rounded, as described above.
The crossover to the insulating phase is expected to take place at $a\sim 3$.
Thus, the criterion $a=3$ can be taken as reasonable estimation of the crossover temperature $T_\mathrm{i}^\mathrm{*}$ to the insulating phase.
This criterion was also used in Ref.~\cite{Kanda_JSPSJ1994}.
As shown in the phase diagram in Fig.~\ref{Fig2}(c), $T_\mathrm{i}^\mathrm{*}$ determined in this way decreases with increasing $E_\mathrm{J}/E_\mathrm{C}$, followed by the quantum phase transition to the superconducting phase at $E_\mathrm{J}/E_\mathrm{C} \sim 0.32$.
This value is close to those derived through the quantum Monte-Carlo simulations ($E_\mathrm{J}/E_\mathrm{C} \sim$~0.4 in Ref.~\cite{Rojas_PRB1996} and $\sim$~0.35 in Ref.~\cite{Capriotti_PRL2003}) but is somewhat different from that of the previous experiment ($E_\mathrm{J}/E_\mathrm{C} \sim$~0.59 in Ref.~\cite{vanderZant_PRB1996}).
At $E_\mathrm{J}/E_\mathrm{C} \ll 1$, $T_\mathrm{i}^\mathrm{*}$ approaches the charge BKT transition temperature $T^\mathrm{BKT}_\mathrm{i}=E_\mathrm{C}/ \pi \varepsilon  k_\mathrm{B}$ expected for the classical case ($E_\mathrm{J}=0$)~\cite{Fazio_PRB1991} ($\varepsilon \approx 1.16$ for a square lattice~\cite{Bobbert_in_book1991}).

We note that Kanda \textit{et al}.\ obtained $T_\mathrm{i}^\mathrm{*}$ by fitting the temperature dependence of $R$ to the formula expected from the charge BKT transition~\cite{Kanda_JSPSJ1994,Kanda_JSPSJ1995-2}.
Our resistance data, however, show a broad bump with a peak at $\sim$~500~mK [Fig.~\ref{Fig2}(a)], which prevents us from performing a similar analysis.
The origin of the bump is unknown.

As shown in the upper panels of Figs.~\ref{I-V_curve}(d)--(f), the temperature dependence of $a$ at low temperatures is stronger than that expected from the theory ($a=2T^\mathrm{BKT}_\mathrm{i}/T +1$) except for $E_\mathrm{J}/E_\mathrm{C}=$~0.28.
This is because another nonlinear effect associated with negative differential conductance caused by the Bloch oscillations sets in at low temperatures as discussed in the next section.

\begin{figure}
	\begin{center}
				\includegraphics[keepaspectratio]{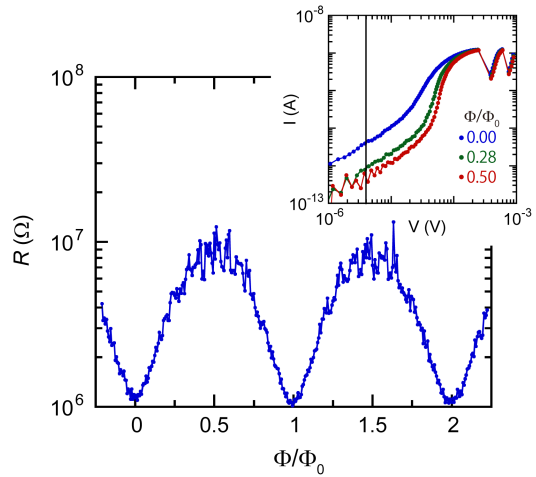}
	\end{center}
	\caption{\label{R_period_Phi_0} 
		Resistance $R$ as a function of magnetic flux $\Phi$ piercing a plaquette of the JJA. The data are taken at 90~mK for $E_\mathrm{J}/E_\mathrm{C}=0.19$, and are measured at a small voltage indicated by the vertical solid line in the inset. Thus, $R$ in the main figure corresponds to the linear part of the $I$--$V$ curve. $R$ has a period of $\Phi_0=h/2e$, indicating that Cooper pairs are responsible for the linear term. The fluctuations in $R$ observed around $\Phi/\Phi_0=$~0.5 and 1.5 are due to noise in current. Inset: $I$--$V$ characteristics taken for three different $\Phi/\Phi_0$. 	}
\end{figure}

The linear coefficient $c^{-1}$ shows the Arrhenius-type temperature dependence $c^{-1} \propto \exp (E_\mathrm{a}/k_\mathrm{B}T)$ at low temperatures as shown in the lower panels of Figs.~\ref{I-V_curve}(d)--(f).
Because $c^{-1}$ corresponds to $R$ taken at small bias voltage, $E_\mathrm{a}$ obtained from $c^{-1} \propto \exp (E_\mathrm{a}/k_\mathrm{B}T)$ should be the same as that obtained from $R$ as a function of $T^{-1}$ displayed in Fig.~\ref{Fig2}(b).
Indeed, $E_\mathrm{a}$ obtained from the two is almost the same as shown in Fig.~\ref{Fig2}(c). 
As discussed above, the linear term $c$ arises from transport of Cooper pairs, not from thermally-excited quasiparticles.
This is also evident from the fact that the linear term has a period of flux quantum $\Phi_0=h/2e$ when magnetic flux piercing a plaquette of JJA, $\Phi$($=BS$), is varied as demonstrated in Fig.~\ref{R_period_Phi_0}, where $B$ is a magnetic field and $S=$~6$\times$6~$\mu$m$^2$ is the area enclosed by a plaquette.
Here, the $\Phi_0$-periodicity reflects the fact that the conduction is carried by particles with charge $-2e$, i.e., Cooper pairs, and arises from the $\Phi_0$-period modulation of the Josephson term in Eq.~(1).
(If quasiparticles are responsible for the linear term, it should have a period of $2\Phi_0$.)
The observation of the $\Phi_0$-periodicity is consistent with the above-described picture that unbound Cooper pairs remaining even below $T_\mathrm{i}^\mathrm{*}$ are responsible for the linear conductance.

The activation energy $E_\mathrm{a}$ derived from $c^{-1} \propto \exp (E_\mathrm{a}/k_\mathrm{B}T)$ corresponds to an energy required for separating a charge pair at a distance up to $\Lambda$.
This energy should be $\sim U_\mathrm{p}(\Lambda)/2$ in the case of $E_\mathrm{J}=0$. 
[For the parameters of our JJAs, $ U_\mathrm{p}(\Lambda)/2 \sim 1.5E_\mathrm{C}$.]
As shown in Fig.~\ref{Fig2}(c), $E_\mathrm{a} \sim 1.6E_\mathrm{C}$ for our smallest $E_\mathrm{J}/E_\mathrm{C}$ ($=$~0.14), being consistent with the expectation.
With increasing $E_\mathrm{J}/E_\mathrm{C}$, $E_\mathrm{a}$ decreases steeply toward zero as $E_\mathrm{J}/E_\mathrm{C}$ approaches the quantum critical point [Fig.~\ref{Fig2}(c)].
This is because $E_\mathrm{J}$ provides a kinetic energy for the charges~\cite{Fazio_PRB1991,Fazio_PhysRep2001}.
Indeed, the kinetic energy provided by $E_\mathrm{J}$ becomes large enough to unbind charge pairs to induce the quantum phase transition to the superconducting phase at the quantum critical point.

In the above discussion, we have not considered excitations with charge $\pm e$.
Although the density of quasiparticles is exponentially small at low temperatures, they could have influences on the formation of the insulating phase. 
Feigel'man \textit{et al.}\ suggested that excited quasiparticles screen the Coulomb interaction between Cooper pairs even if their density is small~\cite{Feigelman_JETPLett1997}.
This effect becomes crucial when the temperature is increased up to around the so-called parity temperature $T^\mathrm{*}$ defined by $k_\mathrm{B} T^\mathrm{*} =\Delta_\mathrm{s} /\ln N_\mathrm{eff}(T^\mathrm{*})$, where  $N_\mathrm{eff}(T)=V\nu(0) \sqrt{8\pi \Delta_\mathrm{s} k_\mathrm{B}T}$ is the effective number of states available for quasiparticles [$V$ is the island volume and $\nu(0)$ is the density of states for the normal metal at the Fermi energy]. 
In our JJAs,  $T^\mathrm{*}$ is estimated to be $\sim$~180~mK.
[We used $V=3.9\times$10$^{-19}$~m$^3$, $\nu (0)=1.45\times$10$^{47}$~m$^{-3}$J$^{-1}$~\cite{book_Kittel2004}, and $\Delta_\mathrm{s}= k_\mathrm{B} \times$2.56~K.] 
We note that $T^\mathrm{*}$ is the temperature at which the number of quasiparticles on an island becomes of the order unity~\cite{Tuominen_PRL1992,Averin_PhysicaB1994}.
The effect of the screening at $T \ll \Delta_\mathrm{s} / k_\mathrm{B}$ can be described in terms of self-capacitance $C_\mathrm{g}^\mathrm{qp}=  (2e)^2 V\nu(0) \sqrt{\frac{2\pi \Delta_\mathrm{s} }{ k_\mathrm{B}T}} e^{-\Delta_\mathrm{s} / k_\mathrm{B}T }$~\cite{Feigelman_JETPLett1997}.
It decreases exponentially with lowering $T$ because of the reduction of the quasiparticle density, and it becomes an order of magnitude smaller than $C_\mathrm{g}$ ($=$~14~aF) at 110~mK ($C_\mathrm{g}^\mathrm{qp} \sim$~5~aF).
Thus, the JJAs with $T_\mathrm{i}^\mathrm{*} \leq$~110~mK (i.e., $E_\mathrm{J}/E_\mathrm{C} \geq$~0.19) are not affected by quasiparticles. 
Only the JJAs with higher $T_\mathrm{i}^\mathrm{*}$ (i.e., smaller $E_\mathrm{J}/E_\mathrm{C}$) could be influenced.
In this region, we have studied two JJAs.
For the JJA with the smallest $E_\mathrm{J}/E_\mathrm{C}$ ($=$~0.14), for example, $C_\mathrm{g}^\mathrm{qp}$ ($\sim$~2~fF) at $T_\mathrm{i}^\mathrm{*}$ ($\sim$~150~mK) is comparable to $C_\mathrm{J}$ ($=$1.7~fF).
Then quasiparticles significantly screen the Coulomb interaction at $T_\mathrm{i}^\mathrm{*}$, which should further smear the BKT transition.

So far, we have neglected influence of random background offset charges which may be induced on the islands due to charges trapped in the substrate or the insulating barriers.
Such offset charges should be distributed between $-e/2$ and $+e/2$ because quasiparticles partly compensate the offset charges.
Delsing \textit{et al}.\ suggested that the influence is small because a Cooper-pair soliton, which covers about a hundred ($\sim \Lambda ^2$)  of islands, sees an offset charge averaged over $\sim \Lambda ^2$ islands~\cite{Delsing_PRB1994}.
Apart from this suggestion, there have been some theoretical studies on influence of offset charges~\cite{Zaikin_CJP1996,Middleton_PRL1993,Granato_PRB1986}.
To our best knowledge, however, none has investigated transport of the insulating phase of a 2D JJA in a realistic situation.
Only Zaikin and Panyukov studied it under a practical situation but in a {\it normal} junction array~\cite{Zaikin_CJP1996}.
They investigated an array with offset charges distributed between $-e/2$ and $+e/2$ having a particular correlation in the case of no screening of the Coulomb interaction by the ground capacitance.
Their result suggests that a finite density of free charges is generated at finite temperatures, which alters the charge BKT transition into the Arrhenius-type behavior with an activation energy of $E_\mathrm{a} \sim 0.18 E_\mathrm{C}$. 
In our JJAs, however, the screening of the interaction by the capacitance to ground already alters the BKT transition into the Arrhenius-type resistance at low temperature, as mentioned above.  
Thus, the problem now is how the Arrhenius-type resistance caused by the screening is modified by the presence of offset charges in JJAs. 
It is also a problem how the nonlinear transport is altered if the offset charges are taken into account.

\section{Bloch oscillations}

The $I$--$V$ characteristics exhibit the back-bending characterized by the negative differential conductance ($dI/dV<0$) at low temperatures  [Figs.~\ref{I-V_curve}(b) and (c)] as mentioned above.
The similar back-bending has been reported in the Coulomb-blockade or insulating regime of current-biased single Josephson junctions~\cite{Kuzmin_PRL1991,Watanabe_PRL2001,Watanabe_PRB2003}, one-dimensional (1D) JJAs~\cite{Haviland_JLTP2000,Shimada_JPSJ2016}, and 2D JJAs \cite{Geerligs_PRL1989}, and is understood by coherent single-Cooper-pair tunneling called Bloch oscillations~\cite{Averin_SovPhysJETP1985,Likharev_JLTP1985,Schon_PhysRep1990,Geigenmuller_PhysicaB1988}.
In Bloch oscillations, as well as in Coulomb blockade and Zener tunneling which will be discussed below, the charging effect of a single charge (a single Cooper pair and a single electron) plays an important role.

\begin{figure}
	\begin{center}
		\includegraphics[keepaspectratio]{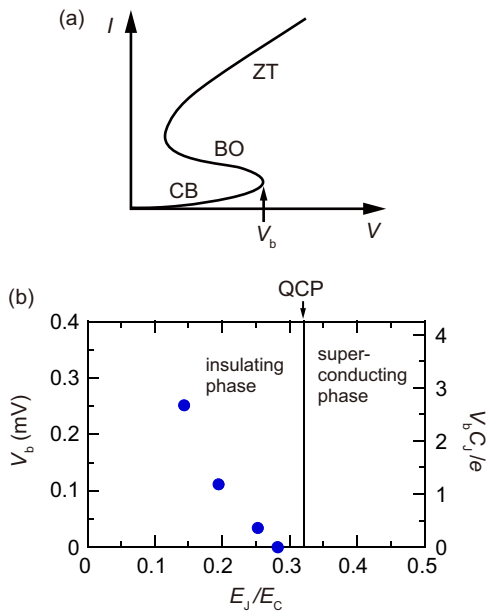}
	\end{center}
	\caption{\label{Bloch_oscillation} 
		(a) Schematic illustration of $I$--$V$ characteristics for a current-biased single Josephson junction with a small $C_\mathrm{J}$. Three regions of Coulomb blockade (CB), Bloch oscillations (BO), and Zener transition (ZT) are indicated. The arrow denotes $V_\mathrm{b}$. (b) $V_\mathrm{b}$ observed in Fig.~\ref{I-V_curve} as a function of $E_\mathrm{J}/E_\mathrm{C}$.
	}
\end{figure}

To obtain basic ideas about Bloch oscillations, we first outline them in the case of a single Josephson junction. 
The Hamiltonian is given by $H=Q^2/2C_\mathrm{J}-E_\mathrm{J}\cos \phi$, where $Q$ and $\phi$ are, respectively, charge on the junction electrode and phase difference across the junction, satisfying $[Q/2e, \phi ]=i$.
Here we consider the charging energy to be comparable to or larger than the Josephson energy.
This Hamiltonian is the same as that of an electron in a periodic lattice, and thus behaviors of the system are understood in terms of Bloch bands with $2e$-periodicity against the quasicharge $q$, which is an analogue of the quasimomentum of an electron in a lattice~\cite{Averin_SovPhysJETP1985,Likharev_JLTP1985,Schon_PhysRep1990,Geigenmuller_PhysicaB1988}.
Such a concept of Bloch bands was recently employed for implementing a novel type of qubit~\cite{Pechenezhskiy_Nature2020}.
Under dc bias current, the single Josephson junction shows the unique $I$--$V$ characteristics as depicted in Fig.~\ref{Bloch_oscillation}(a)~\cite{Averin_SovPhysJETP1985,Likharev_JLTP1985,Schon_PhysRep1990,Geigenmuller_PhysicaB1988}, where stochastic tunneling of single quasiparticles is incorporated to account for the dissipation in the system.  
The $I$--$V$ curve at low external currents is caused by the Coulomb blockade with a small rate of quasiparticle tunneling [Coulomb-blockade (CB) region in Fig.~\ref{Bloch_oscillation}(a)].
At higher currents, Bloch oscillations (BO), which are characterized by the negative differential conductance [BO region in Fig.~\ref{Bloch_oscillation}(a)], set in.
In this region, a coherent single-Cooper-pair tunneling takes place when $q$ reaches the band edge of the Brillouin zone at $q=+e$ and Bragg-reflected to $q=-e$.
The process repeats and voltage across the junction oscillates with a frequency $f=I/(2e)$.
The average voltage across the junction thus decreases, resulting in the negative differential conductance [Fig.~\ref{Bloch_oscillation}(a)].
At high currents, the differential conductance returns positive [Zener-transition (ZT) region in Fig.~\ref{Bloch_oscillation}(a)].
In this region, the system is driven into higher energy bands by Zener transitions, which counterbalances dissipative quasiparticle tunnelings from higher to lower bands.
We note that the structure in the $I$--$V$ curve caused by Coulomb blockade and Bloch oscillations is often called Bloch nose.

Having gained the above picture for a single Josephson junction, we discuss our experimental data on 2D JJAs obtained at low temperatures.
First, we focus on the results of $E_\mathrm{J}/E_\mathrm{C} =$~0.25 [Fig.~\ref{I-V_curve}(b)]. 
With increasing $V$ from zero, $I$ initially shows extremely small values of less than 1$\times$10$^{-13}$ A before $V$ reaches the maximum at $V_\mathrm{b}$ indicated by the arrow (Coulomb-blockade region).
The current in this region is exponentially small [$I \propto \exp (-E_\mathrm{a}/k_\mathrm{B}T)$ with $E_\mathrm{a}\sim E_\mathrm{C}$ for Cooper pairs and with $E_\mathrm{a}\sim \Delta_\mathrm{s}+\frac{1}{4}E_\mathrm{C}$ for quasiparticles] because the energy required to inject for a Cooper-pair soliton is $\sim E_\mathrm{C}$ and that for
a soliton of a thermally-excited quasiparticle into the array is $\sim \frac{1}{4}E_\mathrm{C}$, from the similar discussion in Sect.~\ref{sect:insulating_phase}. 
The voltage then decreases while $I$ keeps increasing, exhibiting negative differential conductance.
In this region, the Bloch oscillations take place.
At higher currents ($I \gtrsim$~1$\times$10$^{-10}$~A), the differential conductance turns positive and the voltage rapidly increases, suggesting that the Zener tunneling occurs. 
Such non-monotonous behaviors are also observed for other JJAs except for the one located closest to the quantum critical point ($E_\mathrm{J}/E_\mathrm{C} =$~0.28).
We note that the signals detected in the negative differential conductance region are noisy, especially for smaller $E_\mathrm{J}/E_\mathrm{C}$, because of the absence of sufficiently large load resistance in the vicinity of the JJA as described in Appendix B in detail.

In single Josephson junctions and also in single normal junctions, phenomena associated with the single charging effect, such as Bloch oscillations and/or Coulomb blockade, are observable only when the junctions are placed in an electromagnetic environment with an impedance much higher than the quantum resistance (i.e., $R_\mathrm{Q}=h/4e^2\approx$~6.45~k$\Omega$ for Cooper pairs and $R_\mathrm{K}=h/e^2\approx$~25.8~k$\Omega$ for electrons)~\cite{Devoret_PRL1990,Girvin_PRL1990,Grabert_ZPhys1991,book_Gravert1992}.
This is because charge fluctuation on the junction capacitance caused by quantum fluctuation of the electromagnetic environment, which otherwise smears the charging effect, is suppressed by isolating the junction from the environment by using a high impedance.
Indeed, a Bloch nose has been observed in single Josephson junctions only when they are placed in a high-impedance environment realized with highly-resistive metal strips or dc-SQUID arrays installed in the vicinity of the junction~\cite{Kuzmin_PRL1991,Watanabe_PRL2001,Watanabe_PRB2003}.
In our experiments for the 2D JJAs, however, the Bloch nose is observed without employing such elaborate wiring, i.e., in the low-impedance environment.
In arrays, charge fluctuation on the junction capacitance is suppressed due to quantization of charge on the islands~\cite{Grabert_ZPhys1991,book_Gravert1992}.
Therefore, when a single charge tunnels a junction, other junctions protect the junction capacitance from charge fluctuation caused by the environment.
Such a protection by other junctions repeats when the charge tunnels junction by junction, allowing for observation of Coulomb blockade and Bloch oscillations in arrays~\cite{Grabert_ZPhys1991,book_Gravert1992}.
We note that, in the low-impedance environment, the tunneling process is considered to be governed by the so-called global rule~
\cite{Grabert_ZPhys1991,book_Gravert1992}, where the tunneling rate is determined by the difference in electrostatic energies of the whole system before and after the tunneling process.

Bloch oscillations in the case of 2D JJAs were theoretically investigated by Sch{\" o}n and Zaikin~\cite{Geigenmuller_Proc1990,Schon_PhysRep1990}.
They studied behaviors of current-biased JJAs by assuming that coherence of Cooper-pair tunneling is maintained over the whole array, which allows the higher-order, simultaneous tunneling of multiple Cooper pairs even at distant junctions.
Their investigation suggests that Bloch oscillations set in when the charge on each outer junction [each of the right and left outermost junctions in Fig.~\ref{Fig1}(a)] reaches $e/M$ in the case of the global rule for quasiparticle tunneling  ($M$ is the number of junctions along the width of the array)~\cite{Geigenmuller_Proc1990,Schon_PhysRep1990}.
The $I$--$V$ characteristics then show a Bloch nose with the maximum voltage $V_\mathrm{b}$ of $\sim \frac{N}{M}\frac{e}{C_\mathrm{J}}$ or a factor smaller.
In our JJAs, the global rule is expected because they are placed in the low-impedance environment as mentioned above~\cite{Grabert_ZPhys1991,book_Gravert1992}.
Therefore, for our JJAs with $N=100$ and $M=100$, $V_\mathrm{b}$ is expected to be $ \sim e/C_\mathrm{J}$ or smaller.
The observed $V_\mathrm{b}$ in our JJAs is of the order of $e/C_\mathrm{J}$ as shown in Fig.~\ref{Bloch_oscillation}(b), which steeply decreases toward zero as $E_\mathrm{J}/E_\mathrm{C}$ approaches the quantum critical point.
The order of magnitude of $V_\mathrm{b}$ is similar to that calculated by Sch{\" o}n and Zaikin, but we cannot reach a clear understanding of the observed behavior because the theoretical study was performed for situations different from our experiments in some respects; 
(i) the current-biased situation (i.e., the high impedance environment) is considered in the theory and (ii) the coherence in Cooper-pair tunneling is assumed to persist across the whole array, which should be destroyed at some length in a real JJA. 
In addition, (iii) the influence of $E_\mathrm{J}$ is not taken into account in the theoretical investigation although our results show the strong suppression of $V_\mathrm{b}$ as $E_\mathrm{J}/E_\mathrm{C}$ approaches the quantum critical point.
Therefore, it urges theoretical investigations for the situation similar to our experiments to reach a quantitative understanding of the observed behavior.

\section{Summary}
Our systematic investigations of nonlinear transport suggest that the crossover from the normal to the insulating phase is understood in terms of the charge BKT transition by including the influences of the finite-range screening of the interaction between Cooper pairs.
The analyses based on the charge BKT transition allows for deducing the crossover temperature to the insulating phase, which continuously decreases toward the quantum critical point.
The transport measurement described here detects averaged nonequilibrium behaviors of excitations.
Further studies of dynamics of individual excitations using, for example, circuit-QED techniques may clarify detailed properties of the crossover to the insulating phase as well as those of the quantum phase transition.

\section*{Acknowledgement}
We acknowledge R. Cosmic, H. Shimada, J. H. Cole, T. Yamamoto, Y. Ashida, and G. Sch{\" o}n for fruitful discussions.
This work was partly supported by JST ERATO (Grant No.~JPMJER1601), JSPS KAKENHI (Grant No.~JP20K03846), and Matsuo Foundation.

\section*{Appendix A: measurement setup}
\begin{figure}
	\begin{center}
		\includegraphics[keepaspectratio]{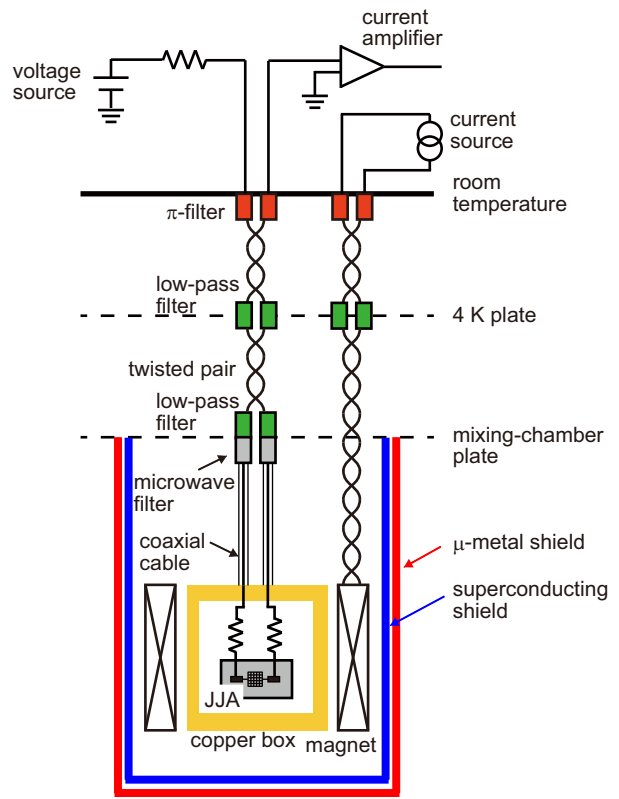}
	\end{center}
	\caption{\label{measurement_setup} 
		Experimental setup of dc transport measurements. A voltage-biasing scheme with a two-probe configuration employed in the insulating phase is shown. The magnetic field is applied perpendicular to the plane of the JJA.
	}
\end{figure}

The experimental setup used for transport measurements is schematically shown in Fig.~\ref{measurement_setup}.
In the insulating phase, a voltage-bias scheme with a two-probe configuration is employed, where the bias voltage $V_\mathrm{bias}$ is applied with a dc source (Yokogawa GS200) and the current $I$ flowing through the JJA is amplified using a current preamplifier (DL Instruments 1211) and detected.
All the wires used for the measurements are carefully filtered. 
Each wire has a $\pi$-filter with a cutoff frequency of 3~MHz at room temperature, RC low-pass filters with a cutoff frequency of 2.4~kHz and 38~kHz at the 4-K and mixing-chamber plates, respectively, and a microwave filter, made of Eccosorb MFS-117 (Emerson \& Cuming), with a cutoff frequency of $\sim$~1~GHz at the base temperature.
A resistance of 19~k$\Omega$ is also inserted in the vicinity of the sample in each wire.
The JJA is installed in a microwave-tight copper box, which is mounted on the mixing-chamber plate of a cryogen-free $^3$He--$^4$He dilution refrigerator (Oxford Instruments Triton200).
The JJA is placed at the center of a superconducting magnet, which is covered with a $\mu$-metal shield and a superconducting shield made of aluminum as shown in Fig.~\ref{measurement_setup}.
Each dc line for the magnet has a $\pi$-filter with a cutoff frequency of 3~MHz at room temperature and an RC low-pass filter with a cutoff frequency of 2.4~kHz at the 4-K plate.

In our voltage-biasing scheme, the voltage on the sample, $V$, is determined from $V=V_{\rm bias} -R_\mathrm{load}I$, where $R_\mathrm{load}$ is the sum of the resistance of electrical wires and the resistance inserted in the wire at room temperature.
We use $R_\mathrm{load}=$~1.2$\times$10$^5$ or 1.12$\times$10$^6$~$\Omega$ depending on the measurements.
Because the resistance of JJAs at low temperatures is orders of magnitude higher than $R_\mathrm{load}$, the current-biased condition is not fully achieved in our setup.

Temperature dependence of $R$ is taken by warming up the mixing chamber slowly at a rate of 60~mK/h at $T \lesssim$~400~mK and at faster rates at higher temperatures. 
Smaller bias voltage is used at lower temperatures to avoid nonlinear effects.
In $I$--$V$ curves measured at fixed temperatures, there is voltage offset of the order of 1~$\mu$V.
The offset is subtracted from $V$ so that the positive and negative voltage parts of the $I$--$V$ curve become symmetric with respect to the origin.
The pulse tube cryocooler used for the dilution refrigerator produces noise in the detected current.
The cryocooler is turned off for a short period (shorter than 30 minutes) when small current ($<$~5$\times$10$^{-11}$~A) is measured.
Such a short stop of the cryocooler does not affect temperature of the mixing chamber.

As mentioned above, the JJA is placed inside the $\mu$-metal shield and the superconducting shield, which reduce the field below 6$\times$10$^{-3}$~Gauss.
The remnant field is further canceled out down to $\sim$~6$\times$10$^{-4}$~Gauss (i.e., $\sim$~1$\times$10$^{-3}\Phi_0$ per plaquette) by applying a magnetic field perpendicular to the plane of the JJA.
The cancellation field is set at the minimum of the resistance observed in the field dependence measured at an appropriate temperature (at 60~mK for $E_\textrm{J}/E_\textrm{C}=$~0.25, for example).

\section*{Appendix B: instability of signals in Bloch-oscillation region}
\setcounter{equation}{0}
\renewcommand{\theequation}{B\arabic{equation}}
As shown in Fig.~\ref{I-V_curve}(c), the signal becomes noisy in the region of the negative differential conductance. 
This is because signals in the negative differential conductance region are unstable in our setup as shown below.

\begin{figure}
	\begin{center}
		\includegraphics[keepaspectratio]{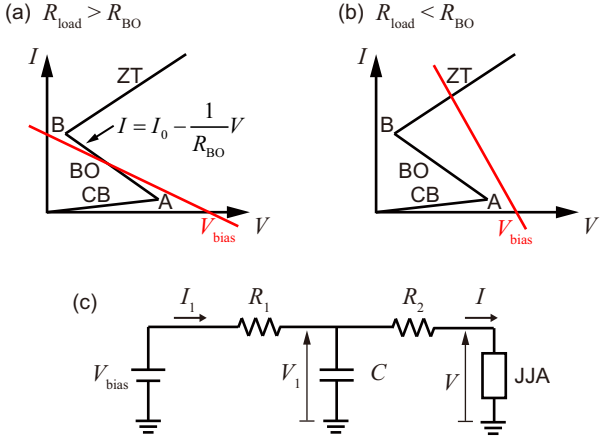}
	\end{center}
	\caption{\label{instability_circuit} 
		(a)(b)~Simplified $I$--$V$ curve of a JJA at low temperatures. The  $I$--$V$ relation in the Bloch-oscillation region is given by Eq.~(\ref{eq:I-V_BO}). The red line indicates Eq.~(\ref{eq:I-V_circuit}). (a)~$ R_\mathrm{load} > R_\mathrm{BO}$ and (b)~$ R_\mathrm{load} < R_\mathrm{BO}$. (c) Simplified model of the measurement circuit. $R_\mathrm{1}$ represents the resistance inserted in the electrical wire at room temperature. $R_\mathrm{2}$ and $C$, respectively, represent total resistance and capacitance of the wire and the filters inside the cryostat.
	}
\end{figure}

We assume that the $I$--$V$ curve of the JJAs is represented by a set of line segments as shown in Fig.~\ref{instability_circuit}(a) for simplicity.
Here, the  $I$--$V$ relation in the Bloch-oscillation region is given, using two constants, $I_\mathrm{0}$ and $R_\mathrm{BO}$, by 
\begin{equation}
I=I_\mathrm{0}-\frac{1}{R_\mathrm{BO}}V ,
\label{eq:I-V_BO}
\end{equation}
where $R_\mathrm{BO}$ corresponds to the negative differential resistance in the Bloch-oscillation region, which is $\sim$~1~M$\Omega$ for the data at 10~mK in Fig.~\ref{I-V_curve}(c).
We also assume that the measurement circuit is modeled by a simple lumped circuit illustrated in Fig.~\ref{instability_circuit}(c).
Here, $R_\mathrm{1}$ is the resistance inserted in the electrical wire at room temperature, and $R_\mathrm{2}$ and $C$, respectively, represent total resistance and capacitance of the wire and the filters inside the cryostat.
In our definition, $R_\mathrm{load}=R_\mathrm{1}+R_\mathrm{2}$.

First, we consider the case of $C=0$.
From Fig.~\ref{instability_circuit}(c), we find 
\begin{equation}
I=\frac{1}{R_\mathrm{load}} \left( V_\mathrm{bias} -V \right) .
\label{eq:I-V_circuit}
\end{equation} 
The current $I$ flowing in the JJA is given by the intersection point between the straight line of Eq.~(\ref{eq:I-V_circuit}) and  the $I$--$V$ curve of the JJA [see Fig.~\ref{instability_circuit}(a)].
Now, we consider that $V_{\rm bias}$ is swept up from zero slowly.
When $V_{\rm bias}$ is small enough, the JJA is in the Coulomb-blockade region, and $I$ is given by the current at the intersection point of Eq.~(\ref{eq:I-V_circuit}) and the $I$--$V$ curve of the Coulomb-blockade region.
When the current reaches point~A, the Bloch oscillations set in.  
If $R_\mathrm{load} > R_\mathrm{BO}$, there is an intersection point on the $I$--$V$ curve of the Bloch-oscillation region.
From Eqs.~(\ref{eq:I-V_BO}) and (\ref{eq:I-V_circuit}), the current flowing though the JJA is given by 
\begin{equation}
I= I_{C=\mathrm{0}} \equiv \frac{V_\mathrm{bias}-R_\mathrm{BO}I_\mathrm{0}}{R_\mathrm{load}-R_\mathrm{BO}}.
\label{eq:I_C=0}
\end{equation} 
With increasing $V_{\rm bias}$ further, the Zener transition sets in at point~B, and the intersection point moves to the $I$--$V$ curve of the Zener transition region.
Now we consider the case of $R_\mathrm{load} < R_\mathrm{BO}$.
In this case, there is no intersection point between Eq. (\ref{eq:I-V_circuit}) and the $I$--$V$ curve of the Bloch-oscillation region after the intersection point passes point~A [see Fig.~\ref{instability_circuit}(b)].
Therefore, the system jumps from the Coulomb blockade to Zener transition regions when $V_{\rm bias}$ is swept up, and the Bloch-oscillation region cannot be traced. 
From the argument in this paragraph, $R_\mathrm{load} > R_\mathrm{BO}$ is required to observe the negative conductance in the Bloch-oscillation region in the case of $C=0$.

Next, we consider the case of $C\ne 0$.
In this case, $I$ and $V$ satisfy the following relations;
\begin{equation}
V_\mathrm{bias} = V_\mathrm{1} + R_\mathrm{1}I_\mathrm{1},
\label{eq:V_bias}
\end{equation} 
\begin{equation}
V_\mathrm{1} = R_\mathrm{2}I + V ,
\label{eq:V_1}
\end{equation} 
and
\begin{equation}
\frac{dV_\mathrm{1}}{dt} = \frac{1}{C}\left(I_\mathrm{1}-I \right) ,
\label{eq:V_1_div}
\end{equation} 
where $V_\mathrm{1}$ and $I_\mathrm{1}$ are the voltage and the current indicated in Fig.~\ref{instability_circuit}(c),
respectively.
In the Bloch-oscillation region, we obtain, from Eqs.~(\ref{eq:I-V_BO}), and (\ref{eq:V_bias})--(\ref{eq:V_1_div}),
\begin{equation}
\frac{dI}{dt} = \frac{1}{\tau} \left(I-I_{C=\mathrm{0}} \right) ,
\label{eq:I_div}
\end{equation} 
where 
\begin{equation}
\tau=\frac{CR_\mathrm{1}\left(R_\mathrm{BO}- R_\mathrm{2} \right)}{R_\mathrm{1}+R_\mathrm{2}-R_\mathrm{BO}}.
\label{eq:tau}
\end{equation} 
Thus, the current is given by 
\begin{equation}
I = I^\mathrm{*} \exp \left(t/\tau \right) +I_{C=\mathrm{0}} ,
\label{eq:I_solution}
\end{equation} 
where $I^\mathrm{*}$ is a constant determined by an initial condition.
If $R_\mathrm{2} < R_\mathrm{BO}$ and $R_\mathrm{1}+R_\mathrm{2} (=R_\mathrm{load}) > R_\mathrm{BO}$, as in our case, $\tau $ is positive, and therefore $I$ does not converge.
Indeed, all the points on the $I$--$V$ curve of the Bloch-oscillation region are unstable points, and $I$ changes as a function of time.
In this case, we observe a different value of $I$ at a different time, and the system gives unstable signals.
For a clear observation of the Bloch-oscillation region,  $\tau <0$ is required in addition to the condition for the case of $C=0$ (i.e., $R_\mathrm{load} > R_\mathrm{BO}$).
These conditions can be satisfied when $R_\mathrm{2} > R_\mathrm{BO}$.
We note that, in the real measurement setup, the capacitance $C$ is distributed over the electrical wires.
This suggests that a resistance sufficiently larger than $R_\mathrm{BO}$ should be inserted in the vicinity of the JJA in each wire for clear observation of the negative differential conductance in the Bloch-oscillation region.


%

\end{document}